\documentclass[11pt]{article}
\usepackage{textcomp}
\usepackage{pdfsync}

\usepackage{amssymb}
\usepackage{amsmath}

\usepackage{graphicx}
\usepackage{latexsym}
\usepackage{appendix}
\usepackage{srcltx}
\def\CA{{\cal A}}
\def\CB{{\cal B}}
\def\CC{{\cal C}}

\def\CE{{\cal E}}

\def\CL{{\cal L}}

\def\CO{{\cal O}}

\def\CS{{\cal S}}

\def\centeron#1#2{{\setbox0=\hbox{#1}\setbox1=\hbox{#2}\ifdim
   \wd1>\wd0\kern.48\wd1\kern-.48\wd0\fi
   \copy0\kern-.48\wd0\kern-.48\wd1\copy1\ifdim\wd0>\wd1
   \kern.48\wd0\kern-.48\wd1\fi}}

\def\JHEP{JHEP~}

\def\PRL{Phys. Rev. Lett.~}
\def\PR {Phys. Rev.~}
\def\CQG {Class. Quant. Grav.~}
\def\PL {Phys. Lett.~}

\newcommand{\beq}{\begin{equation}}
\newcommand{\eeq}{\end{equation}}
\newcommand{\bea}{\begin{eqnarray}}
\newcommand{\eea}{\end{eqnarray}}
\newcommand{\ba}{\begin{array}}
\newcommand{\ea}{\end{array}}

\newcommand{\nn}{\nonumber}

\newcommand{\half}{\frac{1}{2}}

\begin{document}

\hskip3cm

 \hskip12cm{CQUeST--2012-0509}
\vskip3cm

\begin{center}
 \LARGE \bf   Born-Infeld Type Extension of (Non-)Critical Gravity  
\end{center}

\vskip2cm

\centerline{\large Sang-Heon
Yi }

\hskip1cm

\begin{quote}{\large 
Center for Quantum Spacetime, Sogang University, Seoul 121-741,
Korea}
\end{quote}

\vskip1.5cm

\centerline{\bf Abstract} We consider  the Born-Infeld  type extension of (non-)critical gravity which is  higher curvature gravity on Anti de-Sitter space  with   specific combinations of scalar  curvature and Ricci tensor. This theory may also be viewed as  a natural extension of  three-dimensional Born-Infeld new massive gravity to arbitrary dimensions. We show that this  extension  is consistent with holographic $c$-theorem and  scalar graviton modes   are  absent in this theory.  After  showing  that  ghost modes in the theory can be  truncated consistently  by appropriate boundary conditions, we argue that the theory is classically equivalent to Einstein gravity at the non-linear level.  Black hole solutions  are discussed in the view point of the full non-linear classical equivalence between the theory and Einstein gravity.  Holographic entanglement entropy in the theory is also briefly commented on.
  \\
  
\vskip1.5cm
\noindent\underline{\hskip3.5cm}\\
 {\it \small e-mail: shyi(at)sogang.ac.kr}

\thispagestyle{empty}
\renewcommand{\thefootnote}{\arabic{footnote}}
\setcounter{footnote}{0}

\newpage

\section{Introduction}
Holographic principle~\cite{'tHooft:1993gx}  is expected to be deeply related to the fundamental nature of quantum gravity. Though its original insight is based on the semi-classical black hole physics, it is regarded as the fundamental principle of  gravity. Based upon the spectacular progress in string theory, this principle is realized concretely as  the Anti de-Sitter/Conformal Field Theory(AdS/CFT) correspondence~\cite{Maldacena:1997re} which relates $d$-dimensional field theory to   gravity or string theory on $(d+1)$-dimensional AdS space.  According to this correspondence  the large 't Hooft coupling  in field theory corresponds to small curvature scale  in gravity.   This opens up  opportunities to study nonperturbative physics in field theory through gravity.   In this regard, another interesting point in the correspondence is that  the energy scale in field theory corresponds to the extra-dimension in gravity.  This leads to the realization of renormalization group(RG) flow  in the holographic setup, which is far to reach in the traditional perturbative approach. On physical ground it is natural  that   degrees  of freedom are decreased  along   RG flow  since RG flow  is a coarse graining process in quantum field theory. However, it has been known to be very difficult to realize this intuitive picture concretely in continuum field theory except some special cases. One of these special cases is the famous Zamolodchikov $c$-theorem in two dimensional field theory~\cite{Zamolodchikov:1986gt}, which says that the central charge decreases along RG flow.  After the advent of the AdS/CFT correspondence there are various holographic studies about RG flow in higher dimensions, in particular about holographic $c$-theorem and its implication~\cite{Freedman:1999gp}.

In recent years there are some interests in higher derivative gravity with several motivations.  
One motivation in the context of the AdS/CFT  correspondence is to investigate the next order corrections in the large 't Hooft coupling  by higher derivative gravity. In this usual approach, higher derivative terms are treated as small corrections to the Einstein-Hilbert term. This  is the usual strategy  in string theory since stringy corrections are small at low energy. However, if one consider  higher derivative terms as taking  non-small contributions, these are  plagued by ghosts which signal the deficiency of the theory. To avoid ghost problem in this case,  one usually focus on two derivative higher curvature gravity like Gauss-Bonnet or Lovelock gravity.  These higher curvature gravities or their related cousins of   the effective two derivatives nature played important roles in seeing $c$-theorem holographically~\cite{Myers:2010xs}.

The situation is different in three-dimensional gravity  since the usual  Einstein gravity has no propagating degrees of freedom in this case. Through the study of higher derivative terms in three-dimensional topologically massive gravity~\cite{Deser:1982} on AdS space, it has been recognized that the fall-off boundary conditions are important to characterize higher derivative theories. Concretely speaking,   it was found  that by adding higher derivative terms,   the central charge of dual CFT  may vanish at the special value of the coefficient of higher derivative terms. In this case by taking different fall-off boundary conditions in bulk gravity,  boundary dual CFTs are characterized differently.  These  specifications are now    coined as  chiral/log gravity conjectures~\cite{Li:2008dq}. 

Along the line of these developments four-dimensional critical gravity  is proposed as  the analogue of  chiral/log gravity or  critical new massive gravity~\cite{Bergshoeff:2009hq} on three-dimensional AdS space, which contains squared Weyl tensor    as a higher derivative term.  When  the  coefficient of the Weyl tensor term is chosen appropriately,  massless gravitons have   zero excitation energy  and mass of AdS-Schwarzschild black holes  vanishes~\cite{Lu:2011zk}.  Since the linearized   equations of motion for  gravitons in critical gravity takes   the squared form of those in Einstein gravity,   unwanted massive ghost modes are absent. Instead,  there are ghost-like   log modes which fall off more slowly than massless graviton modes  but which may be unseparated  with massless modes.

Another interesting  recent development in  higher curvature gravity is the claim that   four-dimensional conformal gravity on AdS space with appropriate boundary conditions  is equivalent to Einstein gravity on the same space  at the classical but non-linear level~\cite{Maldacena:2011mk}. It was argued that  by  appropriate boundary conditions, ghosts can be truncated consistently or  Einstein solutions are  chosen consistently among all possible solutions. Subsequently, through linear analysis on graviton modes,  non-critical gravity is introduced as the extension of critical gravity by relaxing the condition on the coupling of the Weyl tensor squared term~\cite{Lu:2011ks}. 
In this theory,  when the coupling of higher derivative terms is chosen appropriately,  massive ghost modes fall off more slowly than massless graviton modes  while decoupled from massless modes. Then, massive ghost modes can be truncated consistently by boundary conditions.     

Recently, non-critical gravity is shown to be equivalent to Einstein gravity by using auxiliary field formalism~\cite{Hyun:2011ej}.  To derive this equivalence in the framework of the  AdS/CFT correspondence the Gibbons-Hawking boundary terms as well as the bulk Lagrangian are shown to be reduced to those in Einstein gravity by the consistent choice of subset among all possible solutions in non-critical gravity.  More recently,  four-dimensional non-critical gravity is extended to arbitrary dimensions and the equivalence between this theory and Einstein gravity on AdS space is  verified in various setups at the non-linear level~\cite{Hyun:2012mh}. 

In this paper, we construct the Born-Infeld type  (non-)critical gravity which is a natural  extension of curvature quadratic  (non-)critical gravity. This theory may also be viewed as a natural extension of three-dimensional Born-Infeld new massive gravity~\cite{Gullu:2010} to arbitrary dimensions. 
After its introduction, 
we show that the theory is consistent with holographic $c$-theorem. This consistency with holographic $c$-theorem  may be taken as a beneficial guideline or constraint  for the extension of curvature quadratic (non-)critical gravity   to  even  higher curvature gravity,   as was shown in three-dimensional case~\cite{Sinha:2010}. 
And then, through linear analysis we verify the absence of scalar graviton modes and find the range of the coupling of higher derivative terms to truncate ghosts consistently. Using these results, we argue that this theory is equivalent to Einstein gravity at the non-linear classical level. To see the full non-linear equivalence, black hole solutions and their some properties are also discussed. In the line of this, we briefly comment on the method  to compute  holographic entanglement entropy~\cite{Takayanagi:2011zk} of the theory.

\section{Born-Infeld Gravity}

Non-critical gravity was  introduced as a specific curvature quadratic gravity on AdS space after the truncation of ghosts by appropriate boundary conditions, which is shown to be equivalent to Einstein gravity.   More explicitly, curvature quadratic part  consists only of  scalar curvature square and Ricci tensor square terms with specific coefficients.  It is natural question to ask there are even higher curvature (non-)critical gravity with similar properties. In the following, we consider the critical case simultaneously, in which ghost-like log-modes might be included.   

The hint for the extension of (non-)critical gravity to even higher curvature gravity comes from three-dimensional new massive gravity(NMG) which was extended to even higher curvature gravity theory and named as $R^n$-NMG and BI-NMG~\cite{Gullu:2010}\cite{Sinha:2010}\cite{Nam:2010dd}.   As in the case of three-dimensional NMG, the extension to even higher curvature gravity may be controlled by requiring the  existence of a simple  central charge flow function. However, it is obvious that it becomes harder   to restrict higher curvature couplings in higher dimensions by a single function.    Nevertheless, the consistency with simple holographic $c$-theorem may be taken as a crucial guideline for the extension.

In this section we introduce  the Born-Infeld type extension of    $D$-dimensional   (non-)critical gravity, which we call simply Born-Infeld(BI) gravity.   After the introduction of BI gravity, it is shown to satisfy holographic $c$-theorem, which is the first consistency check for the theory. 
Now,  let us consider the following    $D$-dimensional  action for BI gravity 
\bea  S_{BI} &=&  \frac{1}{2\kappa^2}\int d^Dx \sqrt{-g}\, \CL_{BI} \\
&=& \frac{\sigma}{p(D-2)}\frac{ m^2}{\kappa^2}\int d^Dx \sqrt{-g} \bigg\{ \Big[\det \Big(\delta^{\mu}_{\, \nu} -\frac{1}{m^2}G^{\mu}_{\, \nu} \Big) \Big]^{p} - 1 + \frac{p(D-1)(D-2)^2}{2m^2\ell^2}\bigg\}\,,     \nn  \eea
where $\sigma$ takes $+1$ or $-1$ and $G^{\mu}_{\, \nu}$ denotes  Einstein tensor, $G^{\mu}_{\, \nu} =  R^{\mu}_{\, \nu} -\half R \delta^{\mu}_{\, \nu}$. We take parameters $m^2$ and $\ell^2$ as real valued ones and their range is constrained only by allowance of AdS space. 
In the following  the constant $p$ is taken as 
\beq  p \equiv \frac{1}{D-1} = \frac{1}{d}\,,\eeq
which  turns out to be a  natural  choice  consistent with simple holographic $c$-theorem.  Furthermore, this choice of the value of $p$ leads, expanded in powers of $1/m^2$, to various known (non)-critical gravities as was shown in the following. When $D=3$,  the above action is identical with the one for  BI-NMG up to   some   conventions.

Equations of motion expression,  defined by $\CE_{\mu\nu} \equiv (2\kappa^2/\sqrt{-g}) \delta S_{BI}/\delta g^{\mu\nu}$,   is given by
\bea  - \sigma (D-2)   \CE_{\mu\nu} &=& \frac{m^2}{p}\bigg[   (\det \CA)^p   - 1  + \frac{(D-2)^2}{2m^2\ell^2}\bigg] g_{\mu\nu}    +  2\CB^{\alpha}_{\, (\mu}R_{\nu) \alpha}  -  \CB R_{\mu\nu}         \\
    &&  +  \Big(\nabla_{\alpha}\nabla_{\beta}\CB^{\alpha\beta}  -\nabla^2\CB  \Big)   g_{\mu\nu} +  \nabla_{\mu}\nabla_{\nu}\CB  + \nabla^2 \CB_{\mu\nu}  - 2 \nabla_{\alpha}\nabla_{(\mu}    \CB^{\alpha}_{\nu)}   \,, \nn \eea
where we have defined 
\beq \CA^{\mu}_{\, \nu} \equiv \delta^{\mu}_{\, \nu} - \frac{1}{m^2}G^{\mu}_{\, \nu} \,, \qquad \CB^{\mu}_{\, \nu} \equiv (\det \CA)^{p}(\CA^{-1})^{\mu}_{\, \nu} \,, \qquad \CB \equiv \CB^{\mu}_{\, \mu}\,. \eeq
Our convention  is such that the equations of motion are  read as  $\CE_{\mu\nu} = T_{\mu\nu}$ when matters are coupled with the energy-momentum tensor convention  $T_{\mu\nu} = - (2\kappa^2/\sqrt{-g}) \, \delta S_{mat}/\delta g^{\mu\nu}$.

Whenever the fractional power of the determinant expression is ill-defined,  one should regard it   as a formal expression which may have a concrete meaning after it is expanded in terms of  $G^{\mu}_{\, \nu}/m^2$. 
Using the  Taylor series expansion of the determinant expression, one can see that a few  leading  terms in the action  are given by
\bea S_{BI} &=& \frac{1}{2\kappa^2}\int d^Dx \sqrt{-g}~  \sigma\bigg[ R + \frac{(D-1)(D-2)}{\ell^2} + \CL_2 + \CL_3 + \CO(R^4) \bigg]\,,   \\
&& \CL_{2}  = - \frac{1}{m^2(D-2)} \Big[R^{\mu}_{\,\nu}R^{\nu}_{\, \mu} - \frac{D}{4(D-1)} R^2\Big]\,,    \nn \\ 
  &&      \CL_{3} =   -\frac{2}{3m^4(D-2)}\Big[R^{\mu}_{\,\nu}R^{\nu}_{\, \rho}R^{\rho}_{\, \mu} - \frac{3D}{4(D-1)}RR^{\mu}_{\,\nu}R^{\nu}_{\, \mu}  + \frac{D^2+4D-4}{16(D-1)^2}R^3      \Big]\,.   \nn \eea
Note that the leading Lagrangian is nothing but the one in  Einstein gravity and  the Lagrangian up to $\CL_2$ coincides with the one in noncritical gravity given in~\cite{Hyun:2012mh}\footnote{Interestingly, it was recently discovered that the so-called quasi-topological gravity~\cite{Myers:2010ru} is effectively equivalent to the above Lagrangian up to $R^3$ order when the metric is confined to the conformal flat ones~\cite{dS:2012ef}.}.    

One can easily check that AdS space  is allowed as the vacuum solution  in BI  gravity with 
\beq    \bar{R}_{\mu\nu\rho\sigma} =  - \frac{1}{L^2}  (\bar{g}_{\mu\rho}\bar{g}_{\nu\sigma} - \bar{g}_{\mu\sigma}\bar{g}_{\nu\rho} )\,, \qquad   \bar{R}_{\mu\nu} = - \frac{D-1}{L^2}\bar{g}_{\mu\nu}\,.  \eeq
The radius $L$ of AdS space  is determined in terms of the parameters $m^2$ and $\ell^2$  in the   Lagragian by  the  relation    
\beq  \bigg[1-\frac{(D-1)(D-2)}{2m^2L^2}\bigg]^{\frac{1}{D-1}} \bigg[1-\frac{(D-2)(D-3)}{2m^2L^2} \bigg]=   1-\frac{(D-2)^2}{2m^2\ell^2} \,, \eeq
which comes from the equations of motion $\CE_{\mu\nu}=0$.  Clearly, the relation is the generalization of the three-dimensional BI-NMG case to the $D$-dimensional one. Expanding in powers of $1/m^2L^2$, one obtains
\[\frac{1}{\ell^2} = \frac{1}{L^2}\bigg[1 - \frac{D-4}{4m^2L^2}   - \frac{(D-2)^2(D-6)}{24m^4L^4} + \CO\Big(\frac{1}{m^6L^6}\Big)\bigg]\,, \]
which reproduces the relation among parameters in the corresponding  Lagrangian at each expansion order.

It is very convenient to introduce a useful  quantity $a$,  which appears repeatedly in the following,  as 
\beq  
 a\equiv 1 -\frac{(D-1)(D-2)}{2m^2L^2}\,. \eeq 
Then, $\CA$ and $\CB$ tensors on AdS space are given simply by
\[ \bar{\CA}^{\mu}_{\, \nu} = a\delta^{\mu}_{\, \nu}\,, \qquad  \bar{\CB}^{\mu}_{\, \nu} = a^{pD-1}\delta^{\mu}_{\, \nu} = a^{p}\delta^{\mu}_{\, \nu}\,. \]
Before verifying the consistency of BI gravity with holographic $c$-theorem, let us derive  the so-called A-type trace anomaly, $a^{*}_{d}$  of even $d$-dimensional CFT dual to BI gravity.  One of the easiest way to compute this quantity is the evaluation of the  on-shell Lagrangian on AdS space~\cite{Imbimbo:1999bj}\cite{Myers:2010xs}, which is given in our convention as  
\beq   a^{*}_{d} = - \frac{\pi^{d/2} L^{d+1}}{d\, \Gamma(\frac{d}{2})}\, \frac{1}{2\kappa^2}\,\CL_{BI} \Big|_{AdS} = \frac{ \pi^{d/2}}{\Gamma(\frac{d}{2})} \frac{L^{d-1}}{\kappa^2}\,  \sigma\, a^{p} \,. \eeq
One may note that the final result is simply proportional to the one in Einstein gravity with   factor $\sigma a^p$. This is the first  indication of the   non-linear  equivalence of BI gravity and Einstein gravity with  rescaled Newton's constant, $\kappa^2/\sigma a^{p}$ and substituted cosmological constant, $L$.

In two derivative gravity,  the holographic realization of RG flow between two conformal points in dual $d$-dimensional field theory is given  by  the AdS kink   geometry~\cite{Freedman:1999gp}\cite{Myers:2010xs}. Following this realization,   let us consider the following geometry on $D = (d+1)$ dimensions 
\beq ds^2  = L^2\bigg[dr^2 + e^{2A(r)} \Big(-dt^2 + d{\bf x}^2_{D-2}\Big ) \bigg] \equiv   L^2\Big[dr^2 + g_{ij}dx^{i}dx^{j} \Big]\,. \eeq
Its non-vanishing curvature tensor components   are   given by
\beq 
R^{r}_{~irj} = -\frac{1}{L^2}(A'' + A'^2)g_{ij}\,, \qquad    R^{i}_{~jkl} = - \frac{A'^2}{L^2}(\delta^{i}_{\, k}\, g_{jl} - \delta^{i}_{\, l}\, g_{jk})\,,  \eeq 
%
where the prime denotes the differentiation with respect to the radial coordinate $r$. 
Note that conformal point of dual field theory or  corresponding AdS space  is represented by  $A(r) =r$.  Non-trivial RG flow is realized  by a certain function $A(r)$ connecting two AdS spaces, for which    some additional matter action $S_{mat}$ needs to be introduced.

For computational purpose, introduce functions $\CC(r)$ and $\CC_0(r)$  as
\bea \CC (r) &\equiv & 1 -\frac{(D-1)(D-2)}{2m^2L^2}A'^2  \,, \\    && \nn \\ 
\CC_{0} (r) & \equiv & \Big[1  -\frac{(D-1)(D-2)}{2m^2L^2}A'^2  - \frac{D-2}{m^2L^2} A''\Big] \,, \nn \eea
which become constants on AdS space as
\[\
\CC(r)|_{AdS} = \CC_{0}(r)|_{AdS} = a\,. \]
In terms of these functions,  the   values of $\CA$ tensor,  $\det \CA$ and $\CB$ tensor   for the above metric 
are  given by  
\bea  && 
\CA^{r}_{\, r}  = \CC\,,  \qquad  \qquad   \CA^{i}_{\, j} = \CC_{0} \delta^{i}_{\, j}\,,   \qquad \qquad  (\det \CA )^{p} =  \CC^{p}\CC_{0}\,,    \nn \\    
  &&  
\CB^{r}_{\, r} =  \CC^{p-1}\CC_{0}\,, \qquad \CB^{i}_{\, j} = \CC^{p}\, \delta^{i}_{\, j}\,.    \nn  \eea
One may note that
\[ \CB^{t}_{\, t} - \CB^{r}_{\, r} = (\CC - \CC_{0})\CC^{p-1}=  \frac{D-2}{m^2L^2}\, \CC^{p-1}\, A''\,. \]

Now, let us introduce  a central charge flow function    as 
\beq a_d(r) \equiv  \frac{\pi^{d/2}}{\Gamma\big(\frac{d}{2}\big)\, A'^{d-1}} \frac{L^{d-1}}{\kappa^2}\, \sigma \CC^{p}\,, \eeq
which  would  represent  the  flow of central charge along RG  trajectory of  CFT dual to BI gravity.  The adequacy of this choice of the central charge function  comes from two points. First,  this function on AdS space  coincides with the holographic representation   of the A-type trace anomaly in even $d$-dimensional field theory (and its odd $d$-dimensional counterpart),  which is given in our convention as
\beq a_d(r)|_{AdS} =  \frac{ \pi^{d/2}}{\Gamma(\frac{d}{2})} \frac{L^{d-1}}{\kappa^2}\,  \sigma\, a^{p}  = a^{*}_{d}   \,. \eeq
Second, this function, when expanded  in powers of $1/m^2L^2$,  reduces to the known form in the case of  noncritical gravity~\cite{Hyun:2012mh}
\beq  a_d(r) =  \frac{\pi^{d/2}}{\Gamma\big(\frac{d}{2}\big)\, A'^{d-1}} \frac{L^{d-1}}{\kappa^2}\, \sigma \bigg[ 1 - \frac{d-1}{2m^2L^2}A'^2 +\CO\Big(\frac{1}{m^4L^4}\Big)\bigg]\,.    \eeq

To address holographic $c$-theorem in this setup, one needs to  evaluate $a'(r)_d$ and see its monotonic property.  By a straightforward calculation one can see that  equations of motion with matters gives us  
\beq     T^{t}_{\, t} - T^{r}_{\, r} = \CE^{t}_{\, t} - \CE^{r}_{\, r} =  \frac{\sigma (D-2)}{L^2} \bigg[ 1 - \frac{D(D-3)}{2m^2L^2}A'^2\bigg]\CC^{p-1}\, A''   \,. \eeq
Some comments are in order.
Since BI gravity is higher derivative theory, it is not guaranteed that   $A''$   appears only as the overall factor  in the above expression.   Indeed,  there are various terms in the form of $A'A'''$, $A''^2$, $A'^2A''^2$, {\it etc}  in each expression  of $\CE_{\mu\nu}$.  In  the above combination  of $\CE^{t}_{\, t} $ and $\CE^{r}_{\, r}$,  various such terms are completely canceled and the final  expression   takes very similar form in Einstein gravity. It is interesting to note that  even the combination contains  infinite number of derivatives as can be seen by expanding $\CC^{p-1}$ in powers of $1/m^2L^2$. 

In the end, one can check that  the  monotonic property of this function  comes from the null energy condition on matters, $T^{t}_{\,t} - T^{r}_{\,r} \le 0$: 
\bea  a'_d(r) &=&  - \frac{\pi^{d/2}}{\Gamma\big(\frac{d}{2}\big)\, A'^{d}} \frac{L^{d-1}}{\kappa^2}\,  \sigma (D-2) \bigg[ 1 - \frac{D(D-3)}{2m^2L^2}A'^2\bigg]\CC^{p-1}\, A''    \\
&=& 
- \frac{\pi^{d/2}}{\Gamma\big(\frac{d}{2}\big)\, A'^{d}} \frac{L^{d+1}}{\kappa^2} 
\Big(T^{t}_{\, t} - T^{r}_{\, r} \Big) \ge 0\,,  \nn \eea 
which shows us that BI  gravity is consistent with the simple  form of holographic $c$-theorem. This is one of  nice properties of BI gravity. (See~\cite{Gullu:2010st} for some related discussions in the three-dimensional BI-NMG case).

One may note that the central charge flow function is proportional to the one in Einstein gravity with simple   factor   $\sigma \CC^{p}$.  Since RG flow or central charge function flow  may be regarded as    non-linear characteristics,  this simple proportionality at the level of  the central charge flow function  between BI gravity and Einstein gravity is a strong indication for the equivalence of  two theories at the non-linear level.

\section{Linear Analysis}
 In this  section we perform linear  analysis of BI gravity  to show that potentially dangerous scalar graviton   modes are absent in BI gravity~\cite{Lu:2011zk}, which was one of  important ingredients to define critical gravity in four dimensions. And then, we show that one can truncate massive ghost modes consistently from massless graviton modes by boundary conditions just like non-critical gravity~\cite{Hyun:2011ej}\cite{Hyun:2012mh}.  By supplementing this linear analysis with the consistency with holographic $c$-theorem in the previous section we argue that  BI gravity  is classically  equivalent Einstein gravity at the non-linear level.     
 
The absence of scalar graviton modes is one of minimum requirements for BI gravity to be equivalent to Einstein gravity at low energy. To see the absence of scalar  graviton modes of BI  gravity, let us expand metric around a background metric $\bar{g}_{\mu\nu}$ as 
\beq g_{\mu\nu} = \bar{g}_{\mu\nu} + h_{\mu\nu} \,. \eeq
Then, the linearized Ricci tensor is given by
\[\delta R_{\mu\nu} = \bar{\nabla}_{\rho}\bar{\nabla}_{(\mu} h_{\nu)\alpha} - \half \bar{\nabla}^2 h_{\mu\nu} - \half \bar{\nabla}_{\mu}\bar{\nabla}_{\nu}h\,, \]
where $h$ denotes the contraction with the background  metric $\bar{g}_{\mu\nu}$ and $\bar{\nabla}$ is covariant derivative with respect to $\bar{g}_{\mu\nu}$. {}From now on, the background geometry is taken as AdS space.

Using the diffeomorphism invariance, one may choose the gauge $ \bar{\nabla}^{\nu}h_{\nu\mu} =  \bar{\nabla}_{\mu}h$.  Under this gauge choice 
%
%
the linearized expression of $\CA$  tensor   is  given by
\[ \delta \CA^{\mu}_{\, \nu} = -\frac{1}{m^2}\delta G^{\mu}_{\, \nu} 
=  \frac{1}{2m^2}\, \Big(\bar{\nabla}^2h^{\mu}_{\, \nu} - \bar{\nabla}^{\mu}\bar{\nabla}_{\nu}h \Big) + \frac{1}{m^2L^2}\Big(h^{\mu}_{\, \nu} + \frac{D-3}{2}\delta^{\mu}_{\, \nu}\, h \Big) \,,  \]
and 
the linearized expression of $\CB$ tensor is given by
\bea
\delta \CB^{\mu}_{\, \nu}  &=&  (\det \bar{\CA})^{-p}\Big[ p\, \bar{\CB}^{\alpha}_{\beta}\, \delta \CA^{\beta}_{\, \alpha}\,  \bar{\CB}^{\mu}_{\, \nu} -   \bar{\CB}^{\mu}_{\, \alpha}\, \delta \CA^{\alpha}_{\, \beta}\, \bar{\CB}^{\beta}_{\, \nu}\Big]     \nn \\ 
&=&
    -\frac{a^{p(2-D)}}{m^2}\Big[ \half \Big( \bar{\nabla}^2h^{\mu}_{\, \nu} - \bar{\nabla}^{\mu}\bar{\nabla}_{\nu}h \Big)    + \frac{1}{L^2}\Big(h^{\mu}_{\, \nu} - \half \delta^{\mu}_{\, \nu}\, h\Big) \Big]\,.  \nn  \eea
It is interesting to note that  the linearized expression of $\CB$ tensor on AdS space contains only  two derivatives on $h_{\mu\nu}$, which brings   the linearized equations of motions into  quartic differential equations. This contrasts with the fact that $\CB$ tensor  generically contains infinite number of derivatives at the non-linear level.   

After some calculation, one can obtain the linearized expression of  contracted equations of motion  as
\beq 
0=\delta \CE^{\mu}_{\, \mu} = \frac{(D-1)(D-2)^2}{2L^2}\Big[1-\frac{D(D-3)}{2m^2L^2}\Big]h\,, 
\eeq
which shows us that the scalar graviton modes  are decoupled and transverse traceless  gauge can be taken.  

Since the absence of scalar graviton modes is established, let us turn to ghost modes. In the transverse and traceless gauge, $\bar{\nabla}^{\mu}h_{\mu\nu}=h=0$,  one can show that   the linearized equations of motion, $\delta \CE_{\mu\nu} =0$ are given  in the following quartic form 
\beq \Big(\bar{\nabla}^2  + \frac{2}{L^2}\Big)\Big(\bar{\nabla}^2  + \frac{2}{L^2}-M^2\Big)h_{\mu\nu}=0\,, \qquad M^2 \equiv (D-2)m^2a\,. \eeq
This quartic differential equation shows  us that there exist massless and  massive modes which are decoupled at the linear level.  On general ground one can see that one of them should be ghosts. We take $\sigma$ appropriately such that  massive modes become ghosts. Since the linearization and the expansion in powers of  $1/m^2L^2$ do not commute, the above linear wave equation, when expanded in power of $1/m^2L^2$, does not reduce to the one in curvature quadratic (non-)critical gravity.

Now, we would like to  specify  BI gravity by restricting the parameter $m^2$. Critical BI gravity is defined by the condition $M^2=0$ or $a=0$.  In terms of parameters $m^2$ and $\ell^2$, the critical point is given by
\[
  m^2 = \frac{(D-2)^2}{2\ell^2} = \frac{(D-1) (D-2)}{2L^2} \,, 
\]
which belongs to the regime $m^2, ~\ell^2 >0$.  
 At this critical point  the   quartic differential equation becomes degenerate and then  $\log$ modes appear instead of massive ghost modes. Furthermore, one can see that the A-type trace anomaly, $a^{*}_{d}$  of even $d$-dimensional dual CFT and its cousin in odd dimensions  vanish at this point. All these are the analogues of curvature quadratic critical gravity\footnote{See also~\cite{Deser:2011xc} for critical gravity on $D$-dimensions.} and signal the null content of critical BI gravity  when log modes, which might be mixed with massless modes,  are truncated.

To define consistent non-critical BI gravity, we should truncate unwanted ghosts.  As in curvature quadratic non-critical gravity, massive ghost modes can be truncated consistently by appropriate boundary conditions since massive ghost modes fall off more slowly than massless modes  as far as $M^2 < 0$, while  the  classical instability by tachyons may be avoided when $M^2$ is above  Breitenlohner-Freedman bound~\cite{Breitenlohner:1982jf},   $ M^2 > - (D-1)^2/4L^2$.  Thus,  we take the range of coupling  $m^2$ in the  non-critical case  as  
\[ \frac{(D-1)(2D^2 - 9 D +9)}{4(D-2)L^2} < m^2 < \frac{(D-1)(D-2)}{2L^2}\,. \]

After  ghost modes are truncated,  non-critical BI  gravity describes massless gravitons just like Einstein gravity.    
 This linear equivalence supplemented by the consistency with holographic $c$-theorem  strongly suggests that non-critical BI gravity is classically equivalent to Einstein gravity at the full non-linear level.  In the next section we test this non-linear equivalence through black hole solutions.

\section{Black Hole Solutions and Their Properties}

It is obvious that  AdS-Schwarzschild black holes in Einstein gravity are also solutions in BI gravity since the  scalar curvature and the Ricci tensor of those black holes are same with those of  the  vacuum AdS space.  In the following,  by computing some  non-linear  quantities in these black holes  we check the non-linear equivalence  between BI gravity and Einstein gravity.  We also comment on holographic entanglement entropy in BI gravity.

In our conventions AdS-Schwarzschild black hole solutions are represented by 
\beq ds^2 =  L^2\bigg[ \frac{d\rho^2}{f(\rho)} -  f(\rho)dt^2   +   \rho^2 d\Omega^2_{D-2} \bigg]\,,  \qquad   f(\rho) \equiv \rho^2 +1  - \frac{\omega }{\rho^{D-3}}\,,  \eeq
where the constant $\omega$ is related to the horizon radius $\rho_H$ as $\omega  = \rho^{D-3}_H(\rho^2_H + 1)$. The Hawking temperature of these black holes are given by
\[ T_H = \frac{1}{2\pi  L} \Big[ \rho_H +\frac{D-3}{2\rho_H}(\rho^2_H + 1)\Big] \,. \] 
By    Wald's formula~\cite{Wald:1993nt} it is straightforward to obtain the Beckenstein-Hawking-Wald entropy of these black holes as
\bea  \CS_{BHW} & =&   \frac{ 2\pi  }{(D-2)\kappa^2}~\int_H  d^{D-2}x \sqrt{h} ~ \sigma \Big[\CB^{\mu}_{\, \mu} - \CB^{t}_{\, t} - \CB^{r}_{\, r} \Big]_{\rho = \rho_H}  ~ \\     
&=& \frac{2\pi  A_{H}}{\kappa^2}\, \sigma\,  a^{p}\,,     \qquad  \qquad  \qquad 
  A_H =   (\rho_HL)^{D-2} \Omega_{D-2} \,,     \nn    \eea
where $A_H$ denotes  the horizon area and $\Omega_{D-2} \equiv 2\pi^{d/2}/\Gamma(d/2)$ is the volume of unit $(D-2)$-sphere. 
Note that the black hole entropy can be obtained  by the corresponding one from Einstein gravity simply by rescaling the Newton's constant  $\kappa^2\rightarrow \kappa^2/\sigma a^{p}$ and substituting cosmological constant $\ell \rightarrow L$. 
One can see that this entropy is related to the  A-type trace anomaly $a^{*}_{d}$ as 
\[
 \CS_{BHW} = 4\pi \, \rho_H^{d-1}\,  a^{*}_{d}\,. 
\]
This shows us that the AdS-Schwarzschild black hole entropy  vanishes whenever the central charge does  in the critical case.  
It is also  interesting to note that this relation is identical with the one in Einstein gravity.  Though  the generic  formula for  holographic entanglement entropy(HEE) in higher derivative gravity is unknown~\cite{deBoer:2011wk}, it is natural to propose that the correct way to compute holographic entanglement entropy in BI gravity is just using Wald formula, which was the case in curvature quadratic non-critical gravity~\cite{Kwon:2012tp}\cite{Hyun:2012mh}. If we adopt this prescription for HEE, one can see that the equivalence of BI gravity and Einstein gravity  persists in HEE, too. 

Now, let us consider mass of these black holes. By assuming,  in our case,  the validity of the second  law of black hole thermodynamics and then  accordingly integrating the differential relation, $dM=T_Hd\CS_{BHW}$,    one can see that mass of AdS-Schwarzschild black holes, $M(BI)$, is given by
\[
M(BI) =  \sigma\, a^{p}\, \frac{(D-2)\omega}{2\kappa^2}\Omega_{D-2} = \sigma\, a^{p}\, M(Einstein)\,, \]
where $M(Einstein)$ denotes mass of AdS-Schwarzschild black holes in Einstein gravity.

One may try to check this result on mass of black holes by  the Euclidean action formalism. 
First, one can see that the on-shell   bulk Euclidean action   on AdS-Schwarzschild black holes (or in fact on any Einstein space)  is given by
\[ I^{bk}_{E}(BI)\Big|  = - \frac{D-1}{D-2} \frac{m^2}{\kappa^2}\sigma \int d^Dx \sqrt{g_E}\Big[(\det \CA)^{p} - 1 + \frac{(D-2)^2}{2m^2\ell^2}\Big]\bigg| = \sigma a^p \frac{D-1}{\kappa^2 L^2} \int d^Dx\sqrt{g_E}\,, \]
which shows us that the on-shell   bulk Euclidean action  is proportional to that of Einstein gravity. To compute mass of black holes  in the Euclidean action formalism, one also needs to obtain the on-shell  Gibbons-Hawking-York boundary term which is not yet known  in  BI gravity. Now, let us assume that the one-shell boundary action on AdS-Schwarzschild black holes is  given by the corresponding one in Einstein gravity with the same factor, $\sigma a^{p}$, in the bulk action\footnote{This assumption is not so groundless since this property was shown~\cite{Hyun:2011ej} to be the case  for  curvature quadratic non-critical gravity which coincides with the expanded form of BI gravity up to relevant order.}. Then,   the  on-shell   Euclidean action on AdS-Schwarzschild black holes  in BI gravity, after an appropriate subtraction by counter terms or background AdS space,  is simply proportional to that in Einstein gravity: $I_E(BI) = \sigma a^p I_{E}(Einstein)$.  As a result, one can obtain mass of AdS-Schwarzschild black holes in BI gravity as 
\beq M(BI) = T_H (I_{E} + \CS_{BHW})(BI) = \sigma a^{p} (I_{E} + \CS_{BHW})(Einstein)   =  \sigma\, a^{p}\, M(Einstein)\,. \eeq
Conversely,
this can be interpreted  as the check of our assumption about the on-shell  Gibbons-Hawking-York boundary term or the verification of the relation $I_E(BI) = \sigma a^p I_{E}(Einstein)$,  if mass of black holes is regarded as already determined by  black hole thermodynamics.

All the above results for AdS-Schwarzschild black holes  are    consistent with the claim on the non-linear equivalence between BI gravity and Einstein gravity at the classical level.

\section{Conclusion}

We have constructed BI gravity as  the extension of (non-)critical gravity, which may also be viewed as the extension of BI-NMG. In the framework of the AdS/CFT correspondence, we showed that BI gravity  is consistent with  holographic $c$-theorem.  Through linear analysis we also showed that scalar graviton modes are absent  and massive ghost modes can be truncated consistently by boundary conditions. By combining various ingredients we argued that BI gravity at non-critical point is classically equivalent to Einstein gravity at the non-linear level.  The way to compute HEE in BI gravity is also proposed consistently with this equivalence. 

There are many ways to elaborate on BI gravity. First of all, there are some missing  elements in BI gravity compared to curvature quadratic (non-)critical gravity.  One interesting missing element  is   the Gibbons-Hawking-York boundary term of  BI gravity, which  is crucial  for well-defined variational principle with boundary.  In the context of the AdS/CFT correspondence this is also important for determining the stress tensor of dual CFT.  Anther missing point  is the absence of  the auxiliary field method, which is very  useful to justify the non-linear equivalence more convincingly. 

Though we have showed  that mass of AdS-Schwarzschild black holes in BI gravity  is given by the corresponding one in Einstein gravity up to a  factor, it would be very interesting to obtain mass of AdS-Schwarzschild black holes in BI gravity as conserved charge, for instance, by using Abbott-Deser-Tekin   approach~\cite{Abbott:1981ff}.

Another interesting question about BI gravity is to investigate its relation to counter terms in  $(D+1)$-dimensional gravity as was the case in three-dimensional BI-NMG~\cite{Jatkar:2011ue}. Though there are various undetermined parameters in counter term actions~\cite{Jatkar:2011ue},   these actions seem to be different from  BI gravity action  in this paper except the three dimensional case. In counter term actions,  the absence  of scalar graviton  is not so clear and the consistency with holographic c-theorem   is not guaranteed, either. However, since there are some open possibilities  to determine counter term actions, it would be interesting to study   further the relation between counter terms and BI gravity.

It is also   interesting  to verify or improve our proposal for HEE computation in BI gravity, since there is no generic prescription for HEE computation.  The study on the gauge/fluid correspondence  and R\'{e}nyi entropy in BI gravity  is also  an interesting future direction.  It is also interesting to find some solutions in BI gravity following BI-NMG case~\cite{Gurses:2011fv}~\cite{Ghodsi:2010ev}.

In the long run, it would be very interesting to find some ways to  confirm   the classical equivalence between BI gravity and Einstein gravity and to understand better the meaning of the consistency of a certain gravity theory with holographic $c$-theorem in general.  

\vskip 2cm 

\centerline{ \Large\bf   Acknowledgments}  
This work is  supported by the National
Research Foundation of Korea(NRF) grant funded by the Korea
government(MEST) through the Center for Quantum Spacetime(CQUeST) of
Sogang University with grant number 2005-0049409.  I would like
to thank Seungjoon Hyun, Wooje Jang, Jaehoon Jeong, Yongjoon Kwon, Soonkeon Nam,  and Jong-Dae Park    for useful discussions.

\newpage

\end{document}